\documentclass[a4paper,twocolumn,11pt,unpublished]{quantumarticle}
\pdfoutput=1
\usepackage[utf8]{inputenc}
\usepackage[english]{babel}
\usepackage[T1]{fontenc}
\usepackage{amsmath}

\usepackage{tikz}
\usepackage{lipsum}
\usepackage{graphicx}
\usepackage[colorlinks=true, allcolors=blue]{hyperref}
\usepackage{amssymb,xcolor}
\usepackage{amsbsy}
\usepackage{amsthm}
\usepackage{amsfonts}
\usepackage{epstopdf}
\usepackage{txfonts}
\usepackage[active]{srcltx}
\usepackage{soul}
\usepackage{subcaption}
\usepackage{comment}

\usepackage{lipsum}

\newcommand\blfootnote[1]{%
  \begingroup
  \renewcommand\thefootnote{}\footnote{#1}%
  \addtocounter{footnote}{-1}%
  \endgroup
}

\newcommand{\beq}{\begin{equation}}
\newcommand{\eeq}{\end{equation}}

\newcommand{\bra}[1]{\langle#1|}

\newcommand{\ket}[1]{|#1\rangle}

\newcommand{\ketbra}[3]{|{#1}\rangle_{#2}\langle{#3}|}
\newcommand{\id}{\leavevmode\hbox{\small1\normalsize\kern-.33em1}}

\newcommand{\tr}{\textnormal{Tr}}

\begin{document}

\title{Experimental protocol for qubit–environment entanglement detection}

\author{G. Bizzarri$^{*, } $}
\affiliation{Dipartimento di Scienze, Universit\'a degli Studi Roma Tre, Via della Vasca Navale 84, 00146 Rome, Italy}
\orcid{0009-0000-4105-0192}
%\email{}
%\homepage{}
\author{L. Sansoni$^{*, } $}\blfootnote{$^{*} $ G.B. and L.S. should be considered joint first authors.}
\affiliation{ENEA - Nuclear Department, Via E. Fermi 45, 00100 Frascati, Italy}
\orcid{0000-0002-4445-1036}
\author{E. Stefanutti}
\affiliation{ENEA - Nuclear Department, Via E. Fermi 45, 00100 Frascati, Italy}
\orcid{0000-0002-2833-7049}
\author{I. Gianani}
\affiliation{Dipartimento di Scienze, Universit\'a degli Studi Roma Tre, Via della Vasca Navale 84, 00146 Rome, Italy}
\orcid{0000-0002-0674-767X}
\author{M. Barbieri}
\affiliation{Dipartimento di Scienze, Universit\'a degli Studi Roma Tre, Via della Vasca Navale 84, 00146 Rome, Italy}
\affiliation{Istituto Nazionale di Ottica - CNR, L.go Enrico Fermi 6, 50126, Florence, Italy}
\orcid{0000-0003-2057-9104}
\author{K. Roszak}
\affiliation{FZU - Institute of Physics of the Czech Academy of Sciences, Na Slovance 1999/2, 182 00 Prague, Czech Republic}
\orcid{0000-0002-9955-4331}
\author{A. Chiuri}
\affiliation{ENEA - Nuclear Department, Via E. Fermi 45, 00100 Frascati, Italy}
\orcid{0000-0001-9733-0740}
\maketitle

\begin{abstract}
Decoherence is a manifestation of the coupling of a system with its environment. The resulting loss of information can hamper the functioning of quantum devices, hence the need of understanding its origin and dynamics. Decoherence can stem from entanglement, but it can also be classical in nature. Indeed, methods have been developed to understand whether qubit-environment entanglement (QEE) is actually present in some important classes of quantum channels - pure dephasing. Their practicality resides in the fact they only require accessing the system qubit. In this article we show an implementation of this technique in a photonic quantum channel simulator
via a scheme that has been tailored to the system under study. By controlling the input state of the environment in our simulation, we can check the occurrence of qubit environment entanglement in simple, yet insightful test cases. Our results showcase the usefulness and experimental relevance of the QEE witnessing technique.

\end{abstract}

\section{Introduction\label{sec1}}

Detecting entanglement directly from a quantum state is not possible in general, as entanglement measures are nonlinear functions of the states. The usual solution is to perform quantum state tomography first \cite{james01,christandl12,ahmed21,hu24} followed by calculation of entanglement from the density matrix \cite{neugebauer20,temporao24}. However, if prior information is available, the problem can be vastly simplified.
When the state is known to be reasonably close to a target, then an entanglement witness represents a handy solution~\cite{PhysRevLett.91.227901,PhysRevLett.92.087902,PhysRevLett.124.160503,Zhou19}. It consists of an observable, built from separable measurements, capable of assessing the presence of entanglement. This technique requires fewer measurement settings than full quantum tomography, and has prompted similar investigations applied to channels, for instance to set bounds on their quantum capacity~\cite{PhysRevLett.119.100502,PhysRevLett.123.090503}. 

%The problem is simplified for pure states, for which entanglement is 
%inversely proportional to the purity of the reduced density matrix of one of the possibly entangled systems.
%This means that tomography is still required, but in a much reduced space. For mixed states shortcuts to the 
%qualification of states as entangled or separable are available only for special classes of density matrices,
%such as X-states \cite{mendonca14}.

For pure states, entanglement is inversely proportional to the purity of the reduced state of either entangled subsystems. This implies that one can extract meaningful information addressing a much-reduced space, with a conspicuous advantage in terms of the number of measurement settings. For mixed states, such shortcuts only work for special classes of density matrices, such as X-states~\cite{mendonca14}. One such class of density matrices involves qubit-environment (QE) states obtained via an interaction
that can only lead to pure dephasing of the qubit once the environment is traced out. Such a case is particularly relevant for solid-state qubits: a mismatch between the energy difference of the qubit states and that of the environmental modes (either natural or induced by external parametres, such as the magnetic field), prohibits energy exchange. Remarkably, it is possible to give qualitative~\cite{roszak15i} and quantitative~\cite{roszak20} information on qubit environment entanglement (QEE) that can be generated
via such a pure dephasing interaction. The only requirement is that the initial qubit state must be pure. 
There are no limitations on the initial state of the environment, which can be (and often is) mixed
and can be of an arbitrary dimension. Thus, there is a distinct asymmetry between the two subsystems which lies at the core of the methods of characterizing entanglement for such states. This asymmetry remains for pure dephasing of larger systems, and it is possible to use methods reminiscent of methods used for the treatment of QEE for the study of arbitrary system-environment entanglement \cite{roszak18i}.

%Such density matrices are particularly relevant for solid state qubits, where decoherence is unavoidable, but a mismatch between the energy difference of the qubit states and the energies of environmental modes (natural or induced by a choice of external parameters, such as the magnetic field for spin qubits) prohibits energy exchange between the two subsystems. 

Generation of QEE via pure dephasing evolutions is unambiguously connected with the transfer of information
from the system state to the state of the environment, while qubit decoherence which is not accompanied
by the generation of entanglement does not involve any information transfer \cite{roszak17}. This means that the QE state is qualitatively different for entangling and separable evolutions: at least in principle, entanglement  could be detected by measurements on a single subsystem, the environment~\cite{roszak15i}. As the environment is, almost by definition, hardly accessible, addressing the qubit is the obvious choice. The measurement of QEE acessing only the qubit is feasible, paying the modest prize of a slightly more complicated scheme~\cite{roszak19,rzepkowski20,strzalka21}, in order to transfer information about entanglement from the state of the environment to the state of the qubit.

In this article, we provide an experimental demonstration of QEE detection in an optical setup, based on the methods of Ref.~\cite{roszak19} for indirect detection. This allowed us to merge the idea of only addressing a single subsystem with the use of a witness observable. The enabling element realising decoherence is a Controlled-NOT (CNOT) gate
between the qubit and its environment - itself in the form of a pure or mixed qubit. As the CNOT gate is an entangling operation, one could conjecture any such decoherence process to be entangling: this is actually true when the QE is in a pure state as long
as the CNOT gate leads to decoherence (the initial state of the environmental qubit is not an equal superposition state
of the form $|\pm\rangle=(|0\rangle\pm |1\rangle)/\sqrt{2}$). 
For mixed states, decoherence without entanglement is possible, and even relatively common 
\cite{eisert02,helm09,maziero10,pernice11,lofranco12}. 

The direct application of the scheme of Ref.~\cite{roszak19}, which involves the use of the same interaction between the qubit 
and the environment that leads to decoherence, to probe entanglement, is not sufficient to witness the entangling power of the 
CNOT gate. This is due to the inherent asymmetry of the gate, which has no effect when the control qubit is in the $|0\rangle$
state. To counter this, we take advantage of the control over the interaction which is possible in optical scenarios, and
use the controlled phase gate in the second step. 

We probe two classes of initial states of the environment for their suitability to generate entanglement under the CNOT operation.
The first class consists of pure state superpositions of the $|0\rangle$ and $|1\rangle$ states that differ by the probability 
of obtaining state $|0\rangle$ quantified by a single parameter $\theta$. The second class of states involves corresponding mixed states,
which have no coherence. For both classes of environmental states, we observe qualitatively correct behavior of the entanglement
witness, which properly signifies qubit-environment entanglement. Quantitatively, we observe a discrepancy which is due to the imbalance between the two polarizations
in the experimental setup. The nature of the discrepancy is well understood and it can be taken into account, so that the theory perfectly reproduces experimental curves. 

The paper is organized as follows. In Sec.~\ref{sec2} we discuss qubit-environment entanglement in our setup, introduce the
measurement scheme, and model the measurement in an idealized scenario. In Sec.~\ref{sec3} we describe the experiment, 
as well as present
and discuss the results. Sec.~\ref{sec4} concludes the paper.

\section{Theory\label{sec2}}
We consider a system qubit (Q) interacting with the environment (E). Although we represent both with single-qubit states, they serve different roles and should be treated asymmetrically: for instance, the environment must only dephase the qubit, but the qubit can cause more general decoherence to the environment. The initial state of Q is required to be pure, while for the environment, we can recur to a more general form of its density matrix
\begin{equation}
\label{env}
R_E(0)= \left(
\begin{array}{cc}
c_0 & d  \\
d^{*} & c_1  \\
\end{array}
\right),
\end{equation}
where $c_0$ and $c_1$ are real, while $d$ is a complex number. The condition for Eq.~\eqref{env} to be a density matix is that $|d|^2 \leq c_0 c_1$, which amounts to non-negative definiteness. The initial state of the joint QE arrangement is 
\begin{equation}
	\label{qe00}
	\sigma(0)=\ket{\psi_i}_Q\bra{\psi_i}\ \otimes\ R_E(0),
\end{equation}
with $\ket{\psi_i}_Q=\alpha\ket{0}_Q+\beta\ket{1}_Q$, and $\alpha,\beta\neq 0$. As shown in Refs \cite{roszak15i,roszak18i}, any state obtained using a conditional quantum gate
\begin{equation}
	\label{gate}
	U=|0\rangle_Q\langle 0|\otimes w_{0,E}+|1\rangle_Q\langle 1|\otimes w_{1,E}
\end{equation}
where $w_{i,E},\ i=0,1$ are conditional environmental operators on state (\ref{qe00}), is entangled if and only if $R_{00}\neq R_{11}$. Here the conditional density matrices of the environment are defined as 
\begin{equation}
\label{rii}
R_{ii}=w_{i,E}R_E(0)w_{i,E}^{\dagger}.
\end{equation}	
The analysis of the entangling character of the decoherence process thus requires a comparison of $R_{00}$ and $R_{11}$, that can be carried out only if these are fiducially prepared. This can be achieved by initialising Q in the pointer state  $|i\rangle_Q$ and applying the gate (\ref{gate}), so that the state $R_{ii}$ of the environment is prepared. The joint state remains separable, since pure dephasing cannot affect pointer states, but there is information on the interaction in the density matrix of the environment, and linearity can be employed to inver what could happen in the more general instance.

%An environment can be prepared in state $R_{ii}$. This is done by initializing the qubit in pointer state $|i\rangle$ and applying the gate (\ref{gate}). The resulting state is a product state, since a pure dephasing gate cannot disturb a pointer state, but there is information in the two environmental density matrices, $R_{ii}$, about the entanglement that would be present if gate (\ref{gate}) was applied to state (\ref{qe0}).

%Here both qubit and environment are represented by single qubit states, but they serve different roles in the system and will be treated  asymmetrically. The qubit initial state is required to be pure, but the environment initial state can be  arbitrary. The most general form of the density matrix of the environment is then

This available information can be extracted from the QE system through measurements of decoherence using a gate in the  form (\ref{gate}), possibly with different conditional environmental operators $w_{iE}'$.
The exact choice of operators $w_{iE}'$ in relation to the operators $w_{iE}$, which describe the effect of the original
conditional gate, determines the set of entangled states that can be witnessed by the procedure. Therefore, the initial QE state for the actual measurement to indirectly detect QEE is
\begin{equation}
\label{qe0}
\sigma(0)=\ket{i}_Q\bra{i} \otimes\ R_E(0),
\end{equation}
with $i=0,1$. The QE system undergoes a two-step evolution, which we implement by applying a series of quantum gates, {\it viz}. a CNOT, a Hadamard on Q and a Controlled-Z (CZ) gate.

\subsection*{Step 1: CNOT}
The CNOT gate is used for the environment to decohere the qubit, and its action is defined as  
\begin{equation}
\label{cnot}
U_1=\ket{0}_Q\bra{0}\ \otimes\ \id_E  + \ket{1}_Q\bra{1} \otimes \sigma_{x,E}.
\end{equation}
When applied to the state \eqref{qe0}, this leads to the joint QE state for $i=0$
\begin{equation}
\label{evo_cnot0}
\sigma_0(t)=U\sigma(0)U^{\dagger} = |0\rangle_Q\langle 0|\otimes R_E(0),
\end{equation}
and for $i=1$,
\begin{equation}
\label{evo_cnot1}
\sigma_1(t)=U\sigma(0)U^{\dagger} = |1\rangle_Q\langle 1|\otimes \sigma_{x,E} R_E(0)\sigma_{x,E}.
\end{equation}
Looking at the reduced density matrix of the environment, it follows that $R_{00}=R_E(0)$ and $R_{11}=\sigma_{x,E} R_E(0) \sigma_{x,E}$. Since QEE is proportional to how different these two density matrices are, we can use any measure of their similarity in order to obtain a proper entanglement measure~\cite{roszak20}: here we use the trace norm~\cite{shao15}, which yields particularly simple analytical results
 \begin{equation}
 	\label{distance}
 	T(R_{00},R_{11})=\frac{1}{2} \tr \sqrt{\Delta R^{\dagger} \Delta R} = \sqrt{\Delta c^2 + e^2}.
 \end{equation}
Here $\Delta R=R_{00}-R_{11}$ is the difference between the two conditional density matrices of the environment, $\Delta c=c_0-c_1$ is the difference between the occupations of the initial density matrix of the environment $R_E(0)$, while $e=2\mathrm{Im}(d)$ is twice the imaginary part of the coherence of that 
matrix (\ref{env}). This means that entanglement is generated only if $\Delta c \neq 0$ and/or $\mathrm{Im}(d) \neq 0$. We stress that the trace norm is not a unique choice and other measures, such as the fidelity, can be employed, yielding the same information 
about quantum correlations.

\subsection*{Step 2: Hadamard on Q + CZ}
\begin{figure}[t!]
    \centering
    \includegraphics[width=\columnwidth]{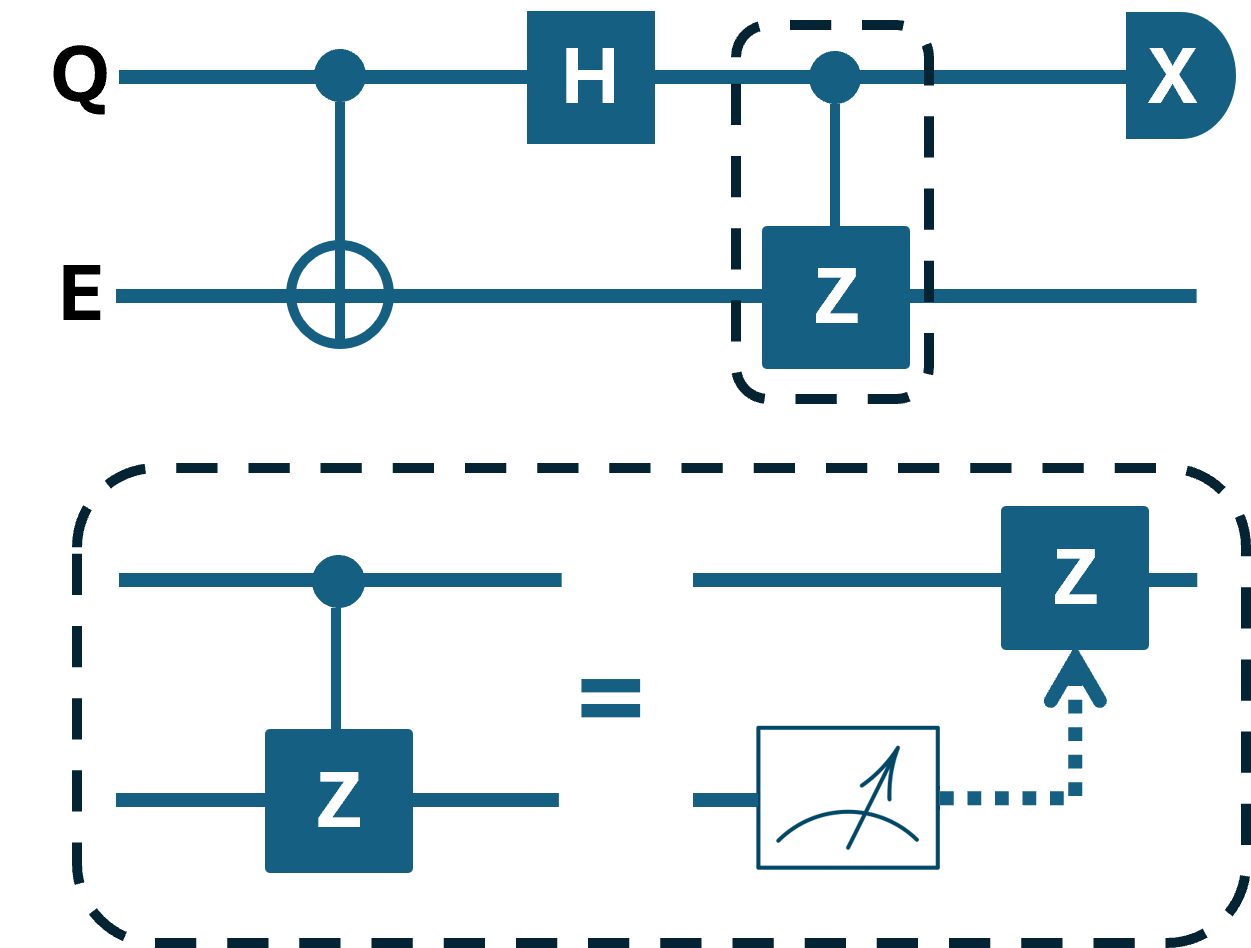}
    \caption{Schematic representation of the experimental protocol for QEE detection.
    Q: qubit, E: environment, H: Hadamard gate, Z (X): $\sigma_Z$ ($\sigma_X$). In the dashed box the feed-forward implementation of a $C-\sigma_Z$ gate used in the experimental protocol as proposed in \cite{PhysRevLett.76.3228} [See the Appendix \ref{appa} for more details].}
    \label{fig:Hadamard_CZ}
\end{figure}

The original scheme of  Ref.~\cite{roszak19} calls for a Hadamard gate on Q in order to map its pointer states to equal superpositions after initial decoherence. This transformation is  realised almost instantaneously and embedded in between free evolutions of the joint QE system. The difference in decoherence experienced after the superposition is excited
depending on the qubit pointer state, 
can serve as a probe for QEE.
The intuition is that, once the environment has been initialised, the qubit should be left to experience the maximal effect from decoherence, making it possible to witness entanglement. The necessary condition for the same QE interaction
to be able to 
witness entanglement as was used to prepare the environmental states is $[w_{0,E},w_{1,E}]\neq 0$. This condition is not fulfilled
by the CNOT gate, since $w_{0,E}=\id$, thus a different controlled operation is required to witness entanglement that is probed.

In our scheme, we take advantage of the control over the QE interaction that is granted by the optical setup,
and use a CZ operation on both qubit and environment after the Hadamard gate is applied. 
A schematic representation of the protocol is shown in Fig.~\ref{fig:Hadamard_CZ}.
The Hadamard gate transforms the QE density matrices (\ref{evo_cnot0}) and (\ref{evo_cnot1}) into
\begin{eqnarray}
\label{h0}
\sigma_{0}^*(t)&=&|+\rangle_Q\langle +|\otimes R_E(0),\\
\label{h1}
\sigma_{1}^*(t)&=&|-\rangle_Q\langle -|\otimes \sigma_{x,E} R_E(0)\sigma_{x,E},
\end{eqnarray}
respectively. Here the qubit states are denoted $|\pm\rangle_Q=\frac{1}{\sqrt{2}}\left(|0\rangle_Q\pm |1\rangle_Q\right)$. The CZ gate,  applied on these new  states, is given by
\begin{equation}
\label{U2}
U_2=\ket{0}_Q\bra{0}\ \otimes\ \id_E \ + \ket{1}_Q\bra{1} \otimes \sigma_{z,E}.
\end{equation}
This can be thought of as resulting from a further interaction within the QE system, or as a part of the measurement process, as we will do in the following.

When acting on the state (\ref{h0}), resulting from the initial qubit state $|0\rangle_Q$, our procedure yields the joint QE
density matrix
\begin{equation}
\label{sigma0}
\sigma'_0 =\frac{1}{2} \left(
\begin{array}{cc}
R_E(0) &  R_E(0) \sigma_{z}  \\
\sigma_{z} R_E(0) & \sigma_{z} R_E(0) \sigma_{z}  \\
\end{array}
\right).
\end{equation}
Note that the QE state (\ref{sigma0}) is written in matrix from only with respect to Q, while the degrees of freedom of the
environment are incorporated in this notation within the matrix elements.
Tracing out the environment yields no change in the occupations of the qubit, since $Tr_{E} R_E(0)=1$, but its coherence does change as 
\begin{equation}
\label{coher0}
    \rho^{(0')}_{01} = \frac{1}{2} \tr_{E} (R_E(0)\sigma_{z})
    =\frac{1}{2}(c_0 - c_1).
\end{equation}
In the second instance, when the initial state of the qubit is $|1\rangle_Q$, the final joint state is,
\begin{equation}
\label{evo_cnot}
\sigma'_1 =\frac{1}{2} \left(
\begin{array}{cc}
\sigma_x R_E(0)\sigma_x &  -\sigma_x R_E(0)\sigma_x \sigma_z  \\
-\sigma_z \sigma_x R_E(0)\sigma_x  & \sigma_{z} \sigma_{x} R_E(0)\sigma_{x} \sigma_{z} \\
\end{array}
\right)
\end{equation}
and the qubit coherence at the end is given by 
\begin{equation}
\label{coher1}
\rho^{(1)'}_{01}= - \frac{1}{2} \tr_{E}(\sigma_x R_E(0)\sigma_x \sigma_z) = \rho^{(0)'}_{01}
\end{equation}
(while the diagonal elements of the density matrix of E remain constant).

In order to detect the presence of entanglement in the QE system, we define an entanglement witness 
following Ref.~\cite{roszak19},
\begin{equation}
    W=\rho^{(0')}_{01}+\rho^{(1')}_{01}=2 \rho^{(0')}_{01}=(c_0-c_1).
    \label{eq:witness_real}
\end{equation}
As expected, the value of this witness depends on the difference between the conditional density matrices $\Delta R$, although the specific action of the controlled-phase gate allows entanglement to be detected only for $c_0\neq c_1$. Entanglement for $c_0 = c_1$ with $e \neq 0$ can be detected using an analogous scheme  utilizing a controlled-$Y$ gate (this scheme is described in the Appendix \ref{App:CY}, \ref{appa}). Notably, $W$ can be assessed by means of a measurement of the $\sigma_x$ Pauli observable on the qubit only. Since we expect $\rho^{(0')}_{01}$ and $\rho^{(1')}_{01}$ to be real, we can directly estimate them as follows:
$\frac{1}{2} \langle \sigma_x \rangle_{\alpha} = \frac{1}{2} Tr[ \sigma_x \rho^{\alpha}] = (\rho^{\alpha})_{01}$ with $\alpha=0,1$.
Thus as the entanglement witness we consider the quantity
\begin{equation}
\label{eq:witness_tot}
    W = \frac{\langle \sigma_x \rangle_{0} + \langle \sigma_x \rangle_{1}}{2}.
\end{equation}

%The plus sign here is the effect of using the Hadamard gate to get the equal-superposition states, which imposes an additional minus sign in the coherence for pointer state $|1\rangle$.

%The superposition is necessary to observe decoherence, while the equal superposition state has maximum coherence and thus enhances the visibility of the effect.

%In order to detect if $R_{00}$ differs from $R_{11}$, a Hadamard gate is performed on the qubit after step 1, yielding superposition states that differ by phase depending on the pointer state $|i\rangle$, so the QE density matrix is given by
%\begin{eqnarray}
%\label{h0}
%\sigma_{0h}(t)&=&|+\rangle\langle +|\otimes R(0),\\
%\label{h1}
%\sigma_{1h}(t)&=&|-\rangle\langle -|\otimes \sigma_x R(0)\sigma_x,
%\end{eqnarray}
%for $i=0,1$ respectively. Here the qubit states are $|\pm\rangle=\frac{1}{\sqrt{2}}\left(|0\rangle\pm |1\rangle\right)$.

\section{Experiment\label{sec3}}
\begin{figure}[t!]
    \centering
    \includegraphics[width=\columnwidth]{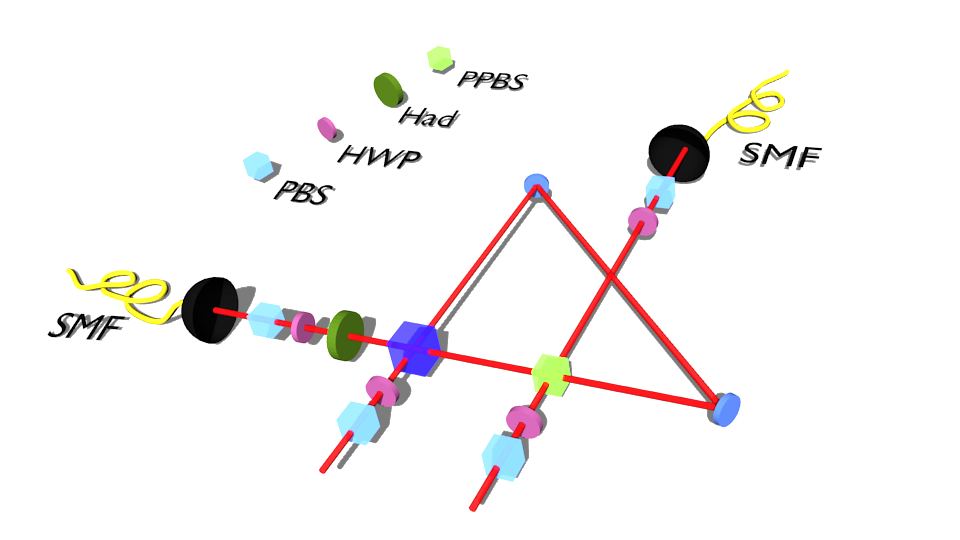}
    \caption{Experimental setup. The two photons representing the system and the environment are prepared in a suitable polarization state by means of a polarizing beam splitter (PBS) and a halfwaveplate (HWP). These two enter the CNOT gate embedded within a Sagnac arrangement with the two paths showing different polarizations. Here interference between the two photons takes place on a PBS, while a partial polarizing beam splitter (PPBS) ensures balancement between the two polarization. A Hadamard gate on the qubit system is applied at the output through a HWP at $22^{\circ}$. The analysis is performed by a tomographic setup composed by a HWP and a PBS. The setup is in free space, but photons are delivered to it by means of single-mode fibers. Likewise, the outputs are fiber-coupled to avalanche photodiodes (APDs) detectors.}
    \label{fig:Setup}
\end{figure}

Our example, however simplified, highlights the essential features of the detection scheme in Ref~\cite{roszak19}, and its modular articulation leads itself to a direct transposition with quantum gates~\cite{PhysRevA.95.042310}. In our implementation, qubit and environment are both implemented in the polarisation of two single photons, with the choice $\ket{0}_Q$ along the horizontal polarisation and $\ket{1}_Q$ along the vertical for the qubit system, while the choice $\ket{0}_E$ along the diagonal polarisation and $\ket{1}_E$ along the antidiagonal polarization for the environment.  The CNOT gate is implemented by means of the setup shown in Fig.~\ref{fig:Setup}.  As reported in Ref.~\cite{cimi20npjqi},  the gate is based on a partially polarising beam splitter within a Sagnac interferometer. While the following Hadamard gate can be implemented by a half-waveplate, the probabilistic functioning of the CNOT gate prevents from putting two of them in cascade~\cite{PhysRevA.75.022313}.  This could be remedied by using initial entanglement~\cite{PhysRevX.5.021010} or, since the second gate is the last operation in the sequence, we can take inspiration from the Griffiths-Niu theorem~\cite{PhysRevLett.76.3228} and realise it by means of measurement and classical feedback~\cite{PhysRevLett.99.250505}. We have taken this second option for the sake of experimental simplicity.

Input states can be controlled by setting  half-waveplates, but, due to additional polarisation-dependent loss from the partially polarising beamsplitters, when the state $\cos 2 \theta\ket{0}_E+\sin 2\theta\ket{1}_E$ is targeted, the angle setting $\alpha$ on the plate should be biased according to $\tan\left( \theta/2\right)=\tan(2\alpha)/\sqrt{3}$ ~\cite{PhysRevLett.95.210506,PhysRevLett.99.250505}. In addition, our gate is effectively a Controlled-Z gate, but it can be mapped onto a proper CNOT operation by using additional Hadamard gates on the environment - these have been included at the preparation and measurement stage in order to keep the setup compact.

\subsection{Environment in a pure state} The initial QE state is given by $\ket{\Psi}_{in}^j=\ket{j}_Q(\cos\frac{\theta}{2}\ket{0}_E+\sin\frac{\theta}{2}\ket{1}_E)$, where $j=0,1$ and $\theta$ is a real parameter that can be experimentally set. After undergoing the transformation described above, the output states are:
\begin{eqnarray}
  \label{eq:Outputstates0}
    \ket{\Psi}_{out}^0=\cos{\frac{\theta}{2}}\ket{+}_Q\ket{0}_E+\sin{\frac{\theta}{2}}\ket{-}_Q\ket{1}_E\\
    \ket{\Psi}_{out}^1=\cos{\frac{\theta}{2}}\ket{+}_Q\ket{1}_E+\sin{\frac{\theta}{2}}\ket{-}_Q\ket{0}_E
    \label{eq:Outputstates1}
\end{eqnarray}
where $\ket{\pm}_Q=\frac{1}{\sqrt{2}}(\ket{0}_Q\pm\ket{1}_Q)$.
Both states are obviously entangled as long as $\cos{\frac{\theta}{2}}\neq 0,1$ and maximum entanglement is reached
for $\cos{\frac{\theta}{2}}=\sin{\frac{\theta}{2}}=\frac{1}{\sqrt{2}}$. As we only target the qubit, we trace out the environment obtaining the following mixed state for both outputs (\ref{eq:Outputstates0}-\ref{eq:Outputstates1}):
\begin{equation}
    \rho_{out}=\frac{1}{2}
    \left(\begin{array}{cc}
        1 & \cos\theta \\
        \cos\theta & 1 
    \end{array}\right).
    \label{eq:rhoout}
\end{equation}
As witness for the entanglement, we use the operator (\ref{eq:witness_tot}), which can be rewritten as
%{\color{blue} \textbf{LS}: the following equation is wrong, as $\rho_{out}$ is already 1-qubit density matrix according to notation of Eq. (\ref{eq:rhoout}), so no partial trace is possible}
\begin{equation}
  W(\theta)=\tr({\rho_{out}\cdot\sigma_x})=\cos\theta,
  \label{eq:witness2}
\end{equation}
for the initial states under study. 
The quantity is equal to zero only for the initial state of the environment that cannot be entangled via the CNOT gate,
with $\theta=\pm\pi/2$. For all 
other values of $\theta$, the pure initial state of E would be entangled with any superposition on Q, which is 
signified by the non-zero value of the witness given by eq.~(\ref{eq:witness2}).

\begin{figure}[t!]
    \centering
    \includegraphics[width=\columnwidth]{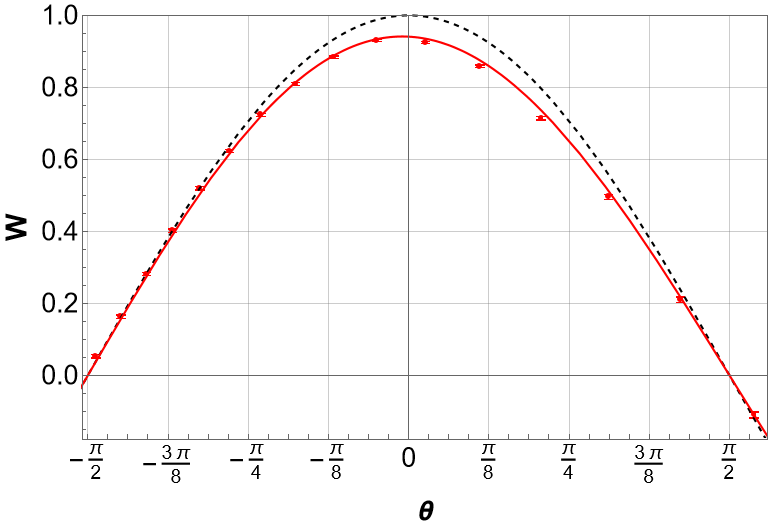}
    \caption{QEE Witness as a function of $\theta$ for pure initial states of the environment. Dots refer to experimental data. The red dashed line represents the theoretical fit of the experimentally obtained data-points, taking into account known experimental imperfections. Black dashed line shows the idealised theoretical behavior.}
    \label{fig:results}
\end{figure}

In the experiment,
for each environmental state corresponding to a specific value of $\theta$, we set the initial qubit state in either $\ket{0}_Q$ or $\ket{1}_Q$ and project the output as explained so far.
The obtained results for the witness (\ref{eq:witness_tot}) as a function of $\theta$ are shown in Fig.~\ref{fig:results}. To account for the expected statistical fluctuations and estimate uncertainties associated with the experimental photon counts, a statistical approach based on a simplified Monte Carlo-like method was adopted. Specifically, the Poisson distribution represents the proper statistical model for describing counting processes in photon detection. Therefore, for each coincidence count $n$ we have generated a set of random variables (i.e.~200) sampled from a Poisson distribution with $n$ as expectation value. This simulated dataset was then used to calculate the standard deviation associated to the corresponding experimental data. 

As can be seen, the experimentally obtained data-points all correspond to non-zero values of the witness, correctly signifying 
QEE for all of the measured values of $\theta$. The curve fitted to these points (red, solid curve) perfectly captures the 
$\theta$-values for which the witness $W(\theta)=0$ and there is no entanglement, namely $\theta=\pm\pi/2$.
The measured and fitted witness curve shows some deviations from the theoretical behavior (plotted using the dashed line
in Fig.~\ref{fig:results}). This deviation between measured and expected behavior is greatest when the values of the witness
are large while it is near zero in the vicinity of the $W(\theta)=0$ points, thus the witness which is meant to only signify the
presence of entanglement and not to act as a measure for it, fullfills its purpouse. 
The deviation is
due to the imbalance between the two polarizations in the experimental setup as well as a non perfect indistinguishability between the qubit and environment photons. Both these contributions can be accounted for. 
In fact, the solid curve used to fit the experimental points is obtained 
when this unavoidable asymmetry is included in the model (see Appendix \ref{app:model} for more details). The data points follow this model with a high accuracy, thus confirming our expected behavior.
%{\color{red}(Please check if I didn't write anything which is not true in this paragraph. Also, I think that we need to explain how the red curve was obtained - in an appendix for instance - because someone is going to ask.)}

% \begin{figure}[t!]
%     \centering
%     \includegraphics[width=\columnwidth]{QEE_pure_X_rescaled.png}
%     \caption{$\langle X \rangle$ as a function of $\theta$. Dots (circles) refer to qubit input $\ket{0}$ ($\ket{1}$). The blue and orange lines represent the theoretical fits.}
%     \label{fig:results}
% \end{figure}

\subsection{Mixed environment} A second experiment has been performed encoding the initial state of the environment as the mixed state
\begin{equation*}
    \rho^{in}_{E}(p) = c_0 \ket{0}_E \bra{0} + c_1 \ket{1}_E \bra{1}
\end{equation*}
with $c_0 \in [0,1]$ and $c_0+c_1=1$. This corresponds to the state (\ref{env}) with the off-diagonal elements set to zero, $d=0$,
and describes a statistical mixture of states, where state $|0\rangle$ has probability $c_0$ and state $|1\rangle$
has probability $c_1$.
According to Eq.~(\ref{eq:witness_real}), the proposed witness reads as follows:
\begin{equation}
\label{wit1}
W(p) = 2c_0-1.
\end{equation}
For such initial states of E, QEE will be generated by the CNOT gate as long as the probabilities corresponding to states $|0\rangle_E$ and
$|1\rangle_E$ differ from one another. 

\begin{figure}[h]
    \centering
    \includegraphics[width=\columnwidth]{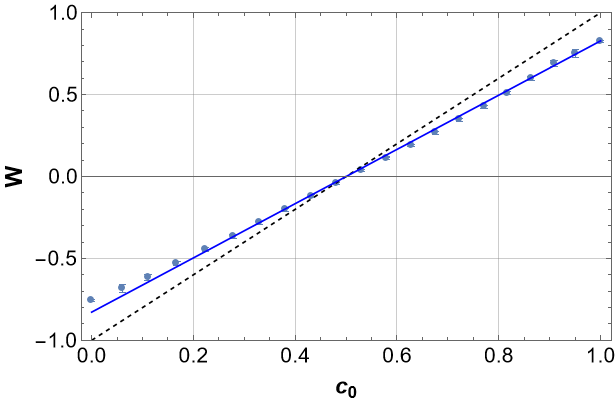}
    \caption{QEE Witness for mixed initial states of the environment as a function of $c_0$. Dots refer to experimental data. The blue solid line represents the theoretical fits considering the experimental parameters. The black dashed line plots the expected theoretical behavior without any imperfection.}
    \label{fig:results_mixed}
\end{figure}

The parameter $c_0$ has been experimentally varied by accumulating the counts obtained for each measurement when the input state for the environment was either $\ket{0}_E$ or $\ket{1}_E$, and  setting different integrating intervals. The proposed witness has been applied to a suitable mixing of these datasets obtained in post-processing.
This procedure has been repeated for both input qubit states $\ket{0}_Q$ and $\ket{1}_Q$ and the results are reported in Fig.~\ref{fig:results_mixed}. 

The experimental data perfectly reproduces the expected linear dependence of the entanglement witness on the probability $c_0$
and correctly signifies the only initial state of the environment incapable of QEE generation under the CNOT operation 
as $c_0=1/2$ (maximally mixed initial state of the environment).
The discrepancy between the expected behavior (dashed black line) of the QEE witness given by eq.~(\ref{wit1})
and the observed behavior (solid blue line) amounts to a difference in slope of the linear $c_0$-dependence 
and does not result in any qualitative changes. Thus the experiment identifies the parameter ranges for the class of mixed 
environmental states that will entangle with a qubit in a superposition state under the CNOT gate, without taking the imbalance between the two polarizations in the experimental setup into account. As in the case of the pure states of the environment
discussed before, this asymmetry can be described theoretically, and the corrected theoretical curves are plotted with a
blue solid line in Fig.~\ref{fig:results_mixed}.

\subsection{Discussion}
It is important to stress here that the presented mixed-state results are the core of the paper. The pure state results serve to 
demonstrate the applicability of the method to different states, but the methods for the study of QEE in pure dephasing
evolutions have been devised for mixed-state environments and their superiority over standard methods of
experimental study of entanglement becomes evident for such states. For pure states, entanglement and decoherence are unambiguously linked and the presence of decoherence signifies the presence
of QEE. Thus quantum state tomography of the QE system is not necessary as long as it is known that the whole QE state is pure - 
information about QEE is present in the density matrix of the qubit.
When considering mixed states, this is no longer the case since the build up of classical correlations can lead to qubit decoherence
just as well as the generation of QEE. 

The mixed state results presented here were generated by mixing results obtained with the environment initially in state
$|0\rangle_E$ and state $|1\rangle_E$ in post-processing. This means that in each run of the experiment, entanglement has been 
signified, and the CNOT gate would generate entanglement if the qubit was initially in a superposition state. This is seen on Fig.~\ref{fig:results_mixed} for $c_0=0$ and $c_0=1$, which both have non-zero values of the witness.
For all intermediate states, the value of the QEE witness is obtained as a mean average of the values at the boundary 
with the probabilities $c_0$ and $c_1$. Thus mixed state results can be obtained from pure state results in a straightforward manner
as in the case of any other observable.

This demonstrates the primary advantage of the presented entanglement witnessing scheme over direct calculation of any 
mixed state entanglement measure, outside of the fact that the operations and measurements are performed only on the qubit. Once the value of the QEE witness for all states in any pure state decomposition of a given mixed state
is known, the witness can be easily found by averaging. For measures that can be applied generally (the scheme only works for pure dephasing).
each mixed state has to be treated separately, since averaging of pure state entanglement is known to lead to wrong results.

This is a distinct advantage even for the two-qubit case and the mixed state which is experimentally studied in this paper is a perfect example. The states obtained via the CNOT gate for the qubit in an equal superposition state and the environment in either state
$|0\rangle_E$ or state $|1\rangle_E$ are both maximally entangled. This means that the average of entanglement of any mixture of these states
is also the maximum value. Nevertheless, it is known that any mixture of the two states is not maximally entangled as long as
$c_0\neq 0, 1$ and for $c_0=1/2$ the state is separable. This information is easily obtained by averaging the values of the
QEE witness over the eigendecomposition of the initial state of the environment, but cannot be found from a straightforward study of the average entanglement in eigendecompositions of the QE state.

% \begin{figure}[t!]
%     \centering
%     \includegraphics[width=\columnwidth]{QEE_Mixed_X.png}
%     \caption{$\langle X \rangle$ as a function of $c_0=p$. Dots (square) refer to experimental data for qubit input $\ket{0}$ ($\ket{1}$). The blue dashed and orange dotted lines represent the theoretical fits considering the experimental parameters. The black lines is the expected behavior without any imperfection.}
%     \label{fig:results_mixed}
% \end{figure}

\section{Conclusions\label{sec4}}
In summary, we demonstrated the implementation of a QEE witness within a quantum photonic simulator.
The procedure has been tailored to a specific experimental platform
and takes advantage from the possibility of tuning the qubit-environment interaction.
We managed to detect correlations using straightforward measurements on just the qubit,
using a series of operations on the qubit in order to transfer the information about entanglement
present in the state of the environment via a QE interaction. 
This success hinges on the straightforward distinction between decoherence that is the result of QEE
and decoherence that is classical in nature, evident in pure dephasing channels. This is at a stark difference from the typical scenario which requires joint measurement of both potentially entangled
systems, and customized approaches targeting specific cases remain important.

The key ability for successful QEE detection consisted of implementing a Hadamard operation with good fidelity and over timescales that are much shorter than the natural timescale of the joint qubit-environment state. This allows to treat the two decoherence processes as separate from 
other operations on the qubit. 
Faulty behavior of other components resulted in deviation of the observed value of the witness with respect to the ideal, but have not compromised its reliability for QEE detection, meaning that
entanglement has been detected correctly and only the magnitude of the observed signal 
has been affected.
All this makes us confident that the QEE witness may earn its place among the everyday toolkit for experimental open quantum systems.

{\it Acknowledgment --}
KR, AC, and ES: This project is funded within the QuantERA II Programme that
has received funding from the EU H2020 research and innovation
programme under GA No 101017733, and with funding organizations MEYS
(Czech Republic) and NQSTI (Italy).
GB is supported by Rome Technopole Innovation Ecosystem (PNRR grant M4-C2-Inv). IG and MB acknowledge support from MUR Dipartimento di Eccellenza 2023-2027 and PRIN project PRIN22-RISQUE-2022T25TR3 of the Italian Ministry of University.

%It is important to note that for pure dephasing evolutions, entanglement manifests itself strongly, and as such is easy to detect. This is a stark difference to the general case, and it allows to multiply the schemes that witness such entanglement and to design schemes tailored especially to the operations and measurements available in a given experimental setup.

\bibliographystyle{quantum.bst}
\bibliography{Bibliography_QEE_Quantum,quantum19}

\onecolumn
\newpage
\appendix
\section{Step 2: Hadamard on Q + CY}\label{App:CY}
We have studied the operation of a protocol for the detection of QEE that can only detect 
entanglement which is the result of a difference in the occupations of the initial state
of the environment. Nevertheless, entanglement will also be present if the initial coherences of E 
have non-zero imaginary parts, as discussed in Sec.~\ref{sec2}. Such entanglement 
can be witnessed if the controlled-phase (CZ) gate in the second step is replaced
by a controlled Y gate, as shown below.

The CY gate, to be applied on the states \eqref{h0} and \eqref{h1}, is given by:
\begin{equation}
U_2'=\ket{0}_Q\bra{0}\ \otimes\ \id_E \ + \ket{1}_Q\bra{1} \otimes \sigma_{y,E}.
\end{equation}
Following the procedure detailed in the main text while replacing the gate $U_2$
of eq.~(\ref{U2}) with the CY gate, we get
coherence terms analogous to \eqref{coher0} and \eqref{coher1} that now read as follows:
\begin{eqnarray}
\rho^{(0')}_{01} &=& \frac{1}{2} Tr_{E} (R_E(0)\sigma_y) = \frac{i}{2} (d- d^{*})= -\frac{e}{2}, \\
\rho^{(1)'}_{01} &=& -\frac{1}{2} Tr_{E} (\sigma_x R_E(0) \sigma_x \sigma_y)  =-\frac{e}{2}=  \rho^{(0')}_{01}.
\end{eqnarray}
%\rho^{(0')}_{01} = \frac{1}{2} Tr_{E} (R(0)\sigma_y) = \frac{1}{2} (\bra{0}R(0) \sigma_y \ket{0}+\bra{1}R(0)\sigma_y \ket{1})=\frac{1}{2} (i \bra{0}R(0) \ket{1}-i \bra{1}R(0)\ket{0})=\frac{1}{2} (i d- i d^{*})=\frac{1}{2} i (d- d^{*})= i \cdot i Im(d) = -Im(d) = -\frac{e}{2}

%\rho^{(1)'}_{01} = -\frac{1}{2} Tr_{E} (\sigma_x R(0) \sigma_x \sigma_y) &= -\frac{1}{2} (\bra{0} \sigma_x R(0) \sigma_x \sigma_y \ket{0}+\bra{1}\sigma_x R(0) \sigma_x \sigma_y \ket{1})= -\frac{1}{2} (i \bra{0}\sigma_x R(0)\sigma_x \ket{1}-i \bra{1} \sigma_x R(0) \sigma_x \ket{0}) &= -\frac{1}{2} (i \bra{1} R(0) \ket{0} -i \bra{0} R(0) \ket{1})=\\ - \frac{1}{2} (i d^{*}- i d)=\frac{1}{2} i (d- d^{*})=  -i Im(d) &= -\frac{e}{2} = -\rho^{(0)'}_{01}

The witness \eqref{eq:witness_real} in this case can detect entanglement for $e \neq 0$,
which can be present in the system also when $c_0 = c_1$:
\begin{equation}
    W=\rho^{(0')}_{01}+\rho^{(1')}_{01}=-2 Im(d)=-e.
    \label{eq:witness_imaginary}
\end{equation}
%This witness will detect only the entanglement which is related with coherences in the initial state of the environment.

\section{Feed-Forward approach\label{appa}}
Here we propose an alternative approach, exploiting two photons with polarization encoding for both the qubit and the environment, relying on a feed-forward method. As stated in \cite{PhysRevLett.76.3228}, whenever the last operation before the measurement stage involves two- (or more) qubit gates, it is possible to map it into single-qubit operations controlled by a classical signal, acting as a feedback.
It is then necessary to design a feed-forward operation equivalent to $C-\sigma_Z$ and $C-\sigma_Y$ operations, in order to retrieve the witnesses \eqref{eq:witness_real} and \eqref{eq:witness_imaginary}.

\subsection*{Feed-forward $C-\sigma_Z$}
The qubit-environment joint state after the interaction and a subsequent rotation of the qubit (see ~\ref{fig:Hadamard_CZ}) is, respectively \eqref{h0} or \eqref{h1}, according to the initial qubit state.

A measurement of the environment (in the computational basis) leads to, respectively,
\begin{eqnarray}
    &\tr_E(\ketbra{0}{_E}{0} \cdot \sigma_0) = c_0\ketbra{+}{_Q}{+},\nonumber\\
    &\tr_E(\ketbra{1}{_E}{1} \cdot \sigma_0) = c_1\ketbra{+}{_Q}{+},\nonumber
\end{eqnarray}
and
\begin{eqnarray}
    &\tr_E(\ketbra{0}{_E}{0} \cdot \sigma_1) = c_1\ketbra{-}{_Q}{-},\nonumber\\
    &\tr_E(\ketbra{1}{_E}{1} \cdot \sigma_1) = c_0\ketbra{-}{_Q}{-}.\nonumber
\end{eqnarray}
Indeed, regardless the initial input state, a measurement of the environment leads to the same qubit state.
In order to distinguish between the two possible outcomes of the environment's measurement, a single-qubit operation on the qubit must be implemented.

By applying a $\sigma_Z$ gate on the qubit whenever $1$ is measured on the environment we retrieve
\begin{equation}
    \begin{cases}
       & \tr_E(\ketbra{0}{_E}{0} \cdot \sigma_0) = c_0\ketbra{+}{_Q}{+}\xrightarrow{Id}c_0\ketbra{+}{_Q}{+},\\
       &\tr_E(\ketbra{1}{_E}{1} \cdot \sigma_0) = c_1\ketbra{+}{_Q}{+}\xrightarrow{\sigma_Z}c_1\ketbra{-}{_Q}{-},
    \end{cases}
    \label{eq:feed_forward_Z_0}
\end{equation}
for $\sigma_0$, and
\begin{equation}
    \begin{cases}
       &\tr_E(\ketbra{0}{_E}{0} \cdot \sigma_1) = c_1\ketbra{-}{_Q}{-}\xrightarrow{Id}c_1\ketbra{-}{_Q}{-},\\
       &\tr_E(\ketbra{1}{_E}{1} \cdot \sigma_1) = c_0\ketbra{-}{_Q}{-}\xrightarrow{\sigma_Z}c_0\ketbra{+}{_Q}{+}.
    \end{cases}
    \label{eq:feed_forward_Z_1}
\end{equation}
By measuring in the $\sigma_Z$ diagonal basis the states \eqref{eq:feed_forward_Z_0} and \eqref{eq:feed_forward_Z_1} the \eqref{eq:witness_real} is retrieved.\\
Therefore, the $C-\sigma_Z$ gate is mapped to a simple $\sigma_Z$ acting on the qubit according to the measure outcome of the environment, in the computational basis. 

\subsection*{Feed-forward $C-\sigma_Y$}

The $C-\sigma_Y$ gate is mapped into a conditional application of a proper single-qubit operation on the initial states \eqref{h0} or \eqref{h1}.
However, in order for the imaginary part of the density matrix to be reconstructed, the environment state has to be mapped into the $\sigma_Y$ diagonal basis via the rotation matrix,
\begin{equation}
    IH = \frac{1}{\sqrt{2}}\begin{pmatrix}
        1 & -i\\
        i & -1
    \end{pmatrix},
    \label{eq:Y_rotation_matrix}
\end{equation}
thus the environment state becomes:
\begin{equation}
    IH R_E(0)IH = \frac{1}{2}\begin{pmatrix}
        & 2i(d-Re(d))+c_0+c_1 \quad -2Re(d)-i(c_0-c_1)\\
        & -2Re(d)-i(c_0-c_1) \quad -2i(d-Re(d))+c_0+c_1\nonumber
    \end{pmatrix}.
\end{equation}

\begin{figure}[t]
\centering
    \includegraphics[width=0.5\textwidth]{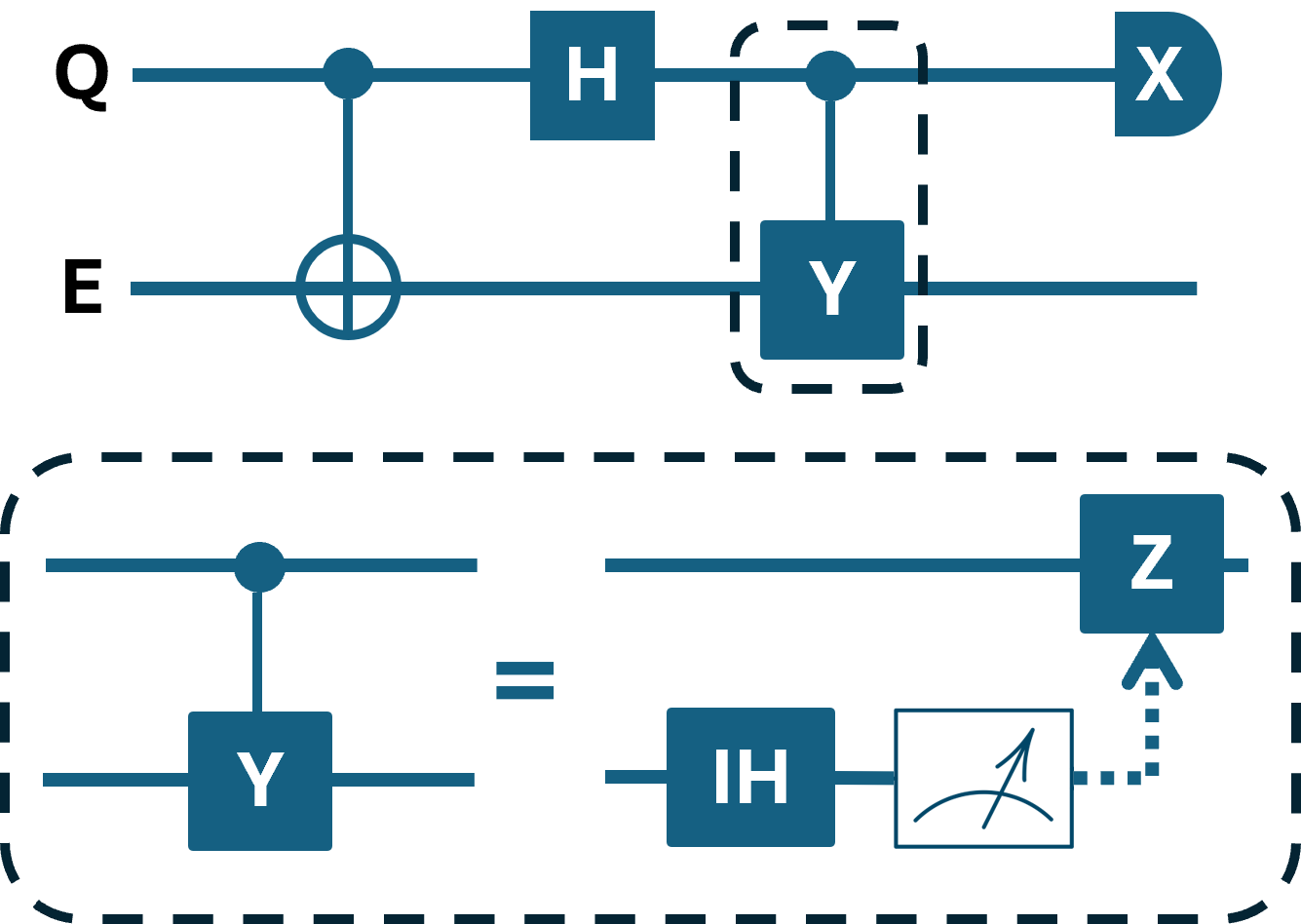}
    \caption{Schematic representation of the experimental protocol for QEE detection. Q: qubit, E: environment, H: Hadamard gate, Z: $\sigma_Z$, X: $\sigma_X$, Y: $\sigma_Y$. In the dashed box the proposed feed-forward implementation of a $C-\sigma_Y$ gate as reported in \cite{PhysRevLett.76.3228}.}
    \label{fig:feed_forward_CY}
\end{figure}

In this case, a projective measurement of the environment in the computational basis leads to
\begin{equation}
    \label{eq:IH_rot_0}
    \begin{cases}
        &\tr_E(\ketbra{0}{E}{0} \cdot \ketbra{+}{_Q}{+}\otimes IH R_E(0)IH) = \left(-Im(d)+\frac{c_0+c_1}{2}\right)\ketbra{+}{_Q}{+},\\
        &\tr_E(\ketbra{1}{E}{1} \cdot \ketbra{+}{_Q}{+}\otimes IH R_E(0)IH) = \left(Im(d)+\frac{c_0+c_1}{2}\right)\ketbra{+}{_Q}{+},
    \end{cases} 
\end{equation}
and we get
\begin{equation}
    IH\cdot \sigma_X R_E(0)\sigma_X \cdot IH = \frac{1}{2}\begin{pmatrix}
        & -2i(d-Re(d))+c_0+c_1 \quad -2Re(d)+i(c_0-c_1)\\
        & -2Re(d)-i(c_0-c_1) \quad 2i(d-Re(d))+c_0+c_1\nonumber
    \end{pmatrix},
\end{equation}
which maps Eq.~\eqref{h1} into
\begin{equation}
    \label{eq:IH_rot_1}
    \begin{cases}
        &\tr_E(\ketbra{0}{E}{0} \cdot \ketbra{-}{_Q}{-}\otimes IH\cdot \sigma_X R_E(0)\sigma_X \cdot IH) = \left(Im(d)+\frac{c_0+c_1}{2}\right)\ketbra{-}{_Q}{-},\\
        &\tr_E(\ketbra{1}{E}{1} \cdot \ketbra{-}{_Q}{-}\otimes IH \cdot \sigma_X R_E(0) \sigma_X \cdot IH) = \left(-Im(d)+\frac{c_0+c_1}{2}\right)\ketbra{-}{_Q}{-}.
    \end{cases} 
\end{equation}
In order to distinguish between the environment's measurement outcome on Eqs~\eqref{eq:IH_rot_0}, \eqref{eq:IH_rot_1}, a $\sigma_Z$ gate is applied on the qubit when $\ketbra{1}{_E}{1}$ is measured, which yields
\begin{equation}
    \begin{cases}
        &\tr_E(\ketbra{0}{E}{0} \cdot \ketbra{+}{_Q}{+}\otimes IH R_E(0) IH)\xrightarrow{Id}\frac{-e+c_0+c_1}{2}\ketbra{+}{_Q}{+},\\
        &\tr_E(\ketbra{1}{_E}{1} \cdot \ketbra{+}{_Q}{+}\otimes IH R_E(0) IH)\xrightarrow{\sigma_X}\frac{+e+c_0+c_1}{2}\ketbra{-}{_Q}{-},
    \end{cases}
    \label{eq:feed_forward_Y0}
\end{equation}
and
\begin{equation}
    \begin{cases}
        &\tr_E(\ketbra{0}{_E}{0} \cdot \ketbra{-}{_Q}{-}\otimes IH\cdot \sigma_X R_E(0)\sigma_X \cdot IH)\xrightarrow{Id}\frac{e+c_0+c_1}{2}\ketbra{-}{_Q}{-},\\
        &\tr_E(\ketbra{1}{_E}{1} \cdot \ketbra{-}{_Q}{-}\otimes IH\cdot \sigma_X R_E(0)\sigma_X \cdot IH)\xrightarrow{\sigma_X}\frac{-e+c_0+c_1}{2}\ketbra{+}{_Q}{+},
    \end{cases}
    \label{eq:feed_forward_Y1}
\end{equation}
respectively,
from which a feed-forward operation equivalent to a $C-\sigma_Y$ gate is implemented. 
A graphical sketch of the operation is presented in Fig.~\ref{fig:feed_forward_CY}.

\section{Modeling of experimental imperfections}
\label{app:model}
In our experiment, we must account for two potential sources of deviation from an ideal setup: the imperfect transmittivities of the partial polarizing beam splitter (PPBS) and the lack of perfect indistinguishability between the qubit and environment photons. The latter affects the visibility of the Hong-Ou-Mandel (HOM) interference, which is crucial for the CNOT operation. The model employed for our setup is derived from the one presented in Ref.~\cite{weinhold2008understanding}.
Let $T_h$ and $T_v$ represent the transmittivities of the PPBS for horizontal and vertical polarizations, respectively, while $V$ denotes the visibility of the HOM interference.
The qubit and environment photons are directed into the two input modes of the PPBS, where they undergo one of the following transformations, depending on their polarizations:
\begin{eqnarray*}
    &&\ket{0}_Q\rightarrow \sqrt{T_h}\ket{0}_Q,\\
    &&\ket{1}_Q\rightarrow \sqrt{T_v}\ket{1}_Q+\sqrt{1-T_v}\ket{1}_E,\\
    &&\ket{+}_E\rightarrow \sqrt{1-\varepsilon^2}\sqrt{T_h}\ket{+}_E+\varepsilon \sqrt{T_h}\ket{+^{\prime}}_E,\\
    &&\ket{-}_E\rightarrow \sqrt{1-\varepsilon^2}(\sqrt{T_v}\ket{-}_Q+\sqrt{1-T_v}\ket{-}_E)+\varepsilon (\sqrt{T_v}\ket{-^{\prime}}_Q+\sqrt{1-T_v}\ket{-^{\prime}}_E),
\end{eqnarray*}
where $\varepsilon=\sqrt{1-\frac{V}{V_{id}}}$, and $V_{id}=0.8$ is the ideal HOM visibility, while the states $\ket{j^{\prime}}$ ($j=\pm$) indicate photons which do not interfere, but still can give some contribution to the measurements.
Note that for the Environment system the horizontal and vertical polarized photons correspond to the diagonal and antidiagonal logic state, respectively.
According to Eq.~\eqref{eq:witness2}, the measurements yield the probability of the QE system being in a particular polarization state. This probability consists of two independent contributions: one from the expected interference between indistinguishable photons  (with weight $\sqrt{1-\varepsilon^2}$) and the other from the spurious contributions of distinguishable states (with weight $\varepsilon$).\\
In our experiment the parameters measured for the transmittivities of the PPBS and the visibility of the HOM interference were $T_h=0.983\pm0.002$, $T_v=0.324 \pm 0.002$ and $V=0.75\pm0.1$.
The models shown in Fig.s \ref{fig:results},\ref{fig:results_mixed} are calculated according to these transformations and measured parameters.

\end{document}